\documentclass[12pt]{article}
\usepackage{epsf, amssymb,amsmath,dsfont}
\usepackage{epsfig}
\usepackage{hyperref}

\makeatletter
\renewcommand\section{\@startsection {section}{1}{\z@}%
                                   {-3.5ex \@plus -1ex \@minus -.2ex}
                                   {2.3ex \@plus.2ex}%
                                   {\normalfont\large\bfseries}}

\renewcommand\subsection{\@startsection{subsection}{2}{\z@}%
                                     {-3.25ex\@plus -1ex \@minus -.2ex}%
                                     {1.5ex \@plus .2ex}%
                                     {\normalfont\bfseries}}

\makeatother

\newcommand{\bea}{\begin{eqnarray}}
\newcommand{\eea}{\end{eqnarray}}
\newcommand{\be}{\begin{equation}}
\newcommand{\ee}{\end{equation}}
\newcommand{\bem}{\begin{pmatrix}}
\newcommand{\eem}{\end{pmatrix}}
\newcommand{\bl}{\begin{align}}
\newcommand{\el}{\end{align}}

\def\a{\alpha}

\def\g{\gamma}

\def\im{\mathrm{Im}}
\def\inf{\infty}

\def\m{\mu}
\def\n{\nu}

\def\o{\omega}  
\def\p{\pi}
\def\pa{\partial}
\def\re{\mathrm{Re}}                
\def\r{\rho}                                     
\def\s{\sigma}                                   
\def\t{\tau}

\def\til{\tilde}

\def\F{\Phi}
\def\G{\Gamma}

\def\L{\Lambda}
\def\O{\Omega}

\def\Q{\Theta}

\def \zz {{\mathbb Z}}

\def \rr {{\mathbb R}}

\begin{document}
\begin{titlepage}

\begin{center}


\hfill ITFA-07-22

\vskip 2 cm {\Large \bf Dying Dyons Don't Count
}
\vskip 1.25 cm  {Miranda C.N. Cheng\footnote{
mcheng@science.uva.nl}, Erik P. Verlinde\footnote{
erikv@science.uva.nl} }\\
{\vskip 0.5cm  Institute for Theoretical Physics\\ University of Amsterdam\\
Valckenierstraat 65\\
1018 XE, Amsterdam\\
The Netherlands\\}

\end{center}

\vskip 1 cm

\begin{abstract}
\baselineskip=16pt

The dyonic \(1\!/4\)-BPS states in 4D string theory with \(\mathcal{N}=4\)
spacetime supersymmetry are counted by a Siegel modular form. The pole structure of the modular form
leads to a contour dependence in the counting formula obscuring its duality invariance.
We exhibit the relation between this ambiguity and the (dis-)appearance
of bound states of  \(1\!/2\)-BPS configurations. Using this insight we propose a precise
moduli-dependent
contour prescription for the counting formula. We then show that the degeneracies are
duality-invariant and are correctly adjusted at the walls of marginal
stability to account for the (dis-)appearance of the two-centered bound
states.
Especially, for large black holes none of these bound states exists at the attractor point and none
of these ambiguous poles contributes to the counting formula. Using this fact we also propose a second,
moduli-independent contour which counts the ``immortal dyons" that are stable everywhere.

\end{abstract}

\end{titlepage}

\pagestyle{plain}
\baselineskip=19pt

\tableofcontents

\vspace{1cm}

\section{Introduction}

A microscopic counting formula was proposed more than a decade ago for dyonic BPS states in the
\({\mathcal{N}}=4, D=4\) theory \cite{Dijkgraaf:1996it} corresponding to the toroidally compactified heterotic string.
The interest for this counting formula was revived about two years ago  in the context of higher-order curvature corrections to the entropy
\cite{LopesCardoso:2004xf}. Subsequently, the formula was given a proper derivation \cite{Shih:2005uc} using the 4D-5D connection \cite{Gaiotto:2005gf}
(See also \cite{Bena:2005ni} for a toroidal example) and the known results for the microscopic counting of five-dimensional black holes \cite{Dijkgraaf:1996xw,Strominger:1996sh}. A particular feature of the formula, namely the occurence of a genus two modular form,  was given a novel interpretation in terms of
string networks \cite{Gaiotto:2005hc}. Recently, a similar counting formula was proposed for a class of more general
\({\mathcal{N}}=4, D=4\) theories \cite{Jatkar:2005bh}, known as the CHL models \cite{CHL}. This class of models
has subsequently been studied from various different angles \cite{CHLdyon}.

In the meantime, various puzzles have been raised about these dyon counting formulas \cite{Sen:2007vb,Dabholkar:2007vk}.
First of all, it has been observed that there is a subtlety in checking their \(S\)-duality invariance.
Secondly, there is an ambiguity in choosing the integration contour arising from the complicated pole structure of
the modular forms that enter the formulas. Finally, it has been noted that the BPS spectrum in the
macroscopic supergravity theory is subjected to moduli dependence due to the presence of walls of marginal stability for some
multi-centered bound states. See \cite{Denef:2007vg} and references therein for a discussion of this phenomenon in the
\({\mathcal{N}}=2\) context.
Finally, either by using a duality argument \cite{Sen:2007vb}, or by studying a specific example in great details \cite{Dabholkar:2007vk},
there have been some hints that all the above issues might actually have something to do with each other.

The goal of the present paper is to address these issues and provide a resolution to some of these puzzles.
In particular, our aim is to present a precise contour prescription that will lead to a counting formula that is
manifestly \(S\)-duality invariant. In fact, we will find two natural prescriptions of this kind, one moduli dependent and a second
only depends on the charges. To arrive at these prescriptions, an important role is played by the one-to-one correspondence between various
poles in the integrand of the counting formula, and the different decay channels in which a dyon can be split
into two \(1\!/2\)-BPS particles. This correspondence between poles and bound states was envisaged in \cite{Dijkgraaf:1996it}, and was
recently reiterated in \cite{Sen:2007vb,Dabholkar:2007vk}.
It turns out that the only poles that can be crossed when the choice of contour is
varied are precisely the ones that admit such a correspondence.
Moreover, we find that the contributions of the poles exactly match the
expected number of states corresponding to the two-centered configurations of
BPS dyons\footnote{Recently this fact was independently noted in
\cite{Sen:2007pg}, which appeared while this paper was being prepared.}.
The key observation which allows us to identify the correct contour prescription is that
the resulting integration contour should render the counting formula explicitly \(S\)-duality invariant, and
should furthermore automatically take the (dis-)appearance of the two-centered bound states into account when a wall of
marginal stability is crossed. This leads to a moduli dependent degeneracy (or index-) formula that counts all the living dyons
in every region of moduli space.
In particular, we observe that the walls of marginal stability have the property that for large black hole charges
(as opposed to ``small black holes" with vanishing leading macroscopic entropy),
none of the two-centered bound states of \(1\!/2\)-BPS particles can exist when the background moduli are fixed at their attractor value.
Using this fact we also propose a second, moduli-independent contour prescription,
which has the property of counting only the ``immortal dyons" which exist everywhere in the moduli space.

The paper is organized as follows. In section two we start by reviewing the dyon counting formula and formulate the issue of its
contour dependence.
The contour dependence of the microscopic counting formula is analyzed in section three. We derive the
condition for a given pole to contribute to the degeneracy formula and calculate its specific contribution.
In section four we give more details of the macroscopic theory
and derive the stability condition for the two-centered bound states of \(1\!/2\)-BPS particles.
 In section five we relate the two sides and present our two contour prescriptions,  one corresponding to the "jumping" index
 and one to the "eternity" index.
Finally, we finish with some discussions and open questions in section six.

\bigskip

\bigskip

\section{The Microscopic Counting Formula and the Poles}
\setcounter{equation}{0}

In this paper we are interested in dyonic BPS states in string compactifications to four dimensions with
\({\mathcal {N}}=4\) space-time supersymmetry. The simplest of these string theories is the \(K3\times T^2
\) compactification of the type II string, or equivalently the toroidally compactificied heterotic string.
Its \(U\)-duality group is the product of the $SL(2,\zz)$ electric-magnetic duality and the \(O(6,22;\zz)\)  $T
$-duality group. A more general class of four-dimensional \({\mathcal {N}}=4\) string theories is obtained by taking
the orbifolds of the aforementioned theory. For definiteness, in the following we will present our
results for the simplest case without any orbifolding, and in the end briefly comment on how, for a class
(the CHL models \cite{CHL}) of the orbifolded theories, the same steps can be modified and followed to
arrive at a very similar result.

The BPS states that preserve one-half of the supersymmetry are well understood.  Using the duality
symmetries, these \(1\!/2\)-BPS states can be mapped to purely electrically charged states corresponding
to heterotic strings carrying only momentum and winding charges. Their degeneracy follows from the level-matching
condition and the state counting of the 24 right-moving
bosonic oscillators of the heterotic string. One finds
\be \label{half_BPS}
d(P) = \oint d\r \, \frac{e^{-i\p P^2 \r}} {\eta^{24}(\r)}\;.
\ee
None of these states carry any macroscopic entropy, at least not at the leading order. To obtain a macroscopic entropy it is necessary to
consider BPS states which preserves only one-quarter of the supersymmetry. Such \(1\!/4\)-BPS states
necessarily carry both electric as well as magnetic charges
\be
(P, Q)\in \G_{6,22}  \oplus  \G_{6,22} \;,
\ee
and their leading macroscopic entropy is given by\footnote{The
short hand notation $|P\wedge Q|$ uses the analogy with the norm of the exterior product of two vectors.
Note the r.h.s. indeed vanishes when $P$ and $Q$ are parallel.}
\be
S=  \p |P\wedge Q| \equiv \p \sqrt{Q^2 P^2 - (Q\cdot P)^2}\;,
\ee
where the inner product is the standard $SO(6,22)$-invariant one on $\Gamma_{6,22}$.
The counting formula for the \(1\!/4\)-BPS states proposed in \cite{Dijkgraaf:1996it} takes the following form:
\be \label{DVV1}
D(P,Q) = \oint_{\mathcal{C}} d\O \,
\frac{e^{-i \p\, \bigl(\begin{smallmatrix} \scriptscriptstyle{P}\\ \scriptscriptstyle{Q}\end{smallmatrix} \bigr)^{\!\dag}\! \O
\bigl(\begin{smallmatrix}  \scriptscriptstyle{P}\\ \scriptscriptstyle{Q}\end{smallmatrix} \bigr)}}{\F(\O) } \,
(-1)^{(P\cdot Q)+1}\, \,,
\ee
where we have incorporated the sign factor $(-1)^{P\cdot Q}$ following \cite{Shih:2005he}.
Here \(\F(\O)\) is an automorphic form of the genus two modular group,
which means that under the  $Sp(2,\zz)$ transformation\footnote{The \(2\times 2\) matrices \(A,\;B,\;C,\;D\) have all integer entries and satisfy the following relation
$$
AB^T = B^T A \;\;\;\;;\;\;\; CD^T = DC^T \;\;\;\;;\;\;\; AD^T- BC^T = {\mathds   {1}}_{2\times2}\;.
$$
}
\be
\O \rightarrow (A\O+B)(C\O+D)^{-1}
\ee
it transforms as
\be
\label{sp2z}
\F(\O) \rightarrow \Bigl(\mbox{det}(C\O+D)\Bigr)^{k}\F(\O)\;.
\ee
For the case of the toroidally compactified heterotic string the weight $k$ is equal to 10.
The automorphic form $\Phi(\O)$ is a well-defined function on the Siegel upper-half plane defined by
\be
\label{upper_half_plane}
\text{det}(\im\O)>0, \qquad\quad \text{Tr}(\im\Omega) >0\;.
\ee
The precise expression for \(\F(\O)\) will not be important for the purpose of our paper.
The main property of $\Phi$ that will concern us is that
it has double zeroes at specific loci in the Siegel domain. In the entropy formula these lead to double poles in the integrand.
When one identifies $\Omega$ with the period matrix of a genus two surface, the poles in $1/\Phi$ occur
precisely at those values of $\Omega$ at which the genus two surface degenerates into two separate genus one surfaces through the pinching of
a trivial homology cycle. These degenerations are labelled by elements of $Sp(2,\zz)$ and
are characterized by the fact that the transformed period matrix is diagonal.
We will write this condition as
\be
\label{SPpoles}
\Bigl( (A\O+B)(C\O+D)^{-1}\Bigr )^\sharp=0\;,
\ee
where the superscript $\sharp$ denotes the upper right (or equivalently, lower left) element of the symmetric $2\times 2$
matrix.

The poles with $C\neq 0$ play an important role in
establishing the correspondence between the macroscopic and microscopic entropy.
However, in this paper we will not be concerned with these entropy-carrying poles, instead we will restrict
our attention mostly to the ones labelled by the elements of $Sp(2,\zz)$ with $C=B=0$ and $A=(D^T)^{-1}$.
These elements constitute the $SL(2,\zz)$ subgroup of  $Sp(2,\zz)$
corresponding to the electric-magnetic or \(S\)-duality.
Under the action of the element
\be
\gamma=\left(\begin{array}{cc}a & b\\ c & d\end{array}\right),\qquad ad-bc=1\;,
\ee
the charges $P$ and $Q$  transform into
\be
\label{sl2z_charge}
\left(\begin{array}{c} P_\g\\Q_\g\end{array}\right)=\g
\left(\begin{array}{c} P\\Q\end{array}\right)\;.
\ee
The period matrix should transform in such a way that the exponent in the counting formula remains invariant, that is
$\Omega \to \Omega_\gamma$ with
\be
\label{transform_O}
\O_\g = \left(\gamma^T\right)^{-1}\!\Omega\gamma^{-1}  .
\ee
Hence, the poles of the Siegel modular form $\Phi(\Omega)$ corresponding to these $SL(2,\zz)$ elements are located at
\be
\label{SLpoles}
\O_\g^\sharp=0\;.
\ee
Due to the presence of these poles, one has to be careful with choosing the contour \(\mathcal{C}\): the counting formula will ``jump'' when the contour crosses
one of these poles. Therefore, strictly speaking the formula (\ref{DVV1}) for $D(P,Q)$ is not just a function of the
charges $P$ and $Q$ but also depends on the contour.

One of the problems with this contour dependence is that it obscures the invariance of the counting formula under the \(S\)-duality.
As mentioned before, the exponential factor in the degeneracy formula (\ref{DVV1}) is invariant under the simultaneaous $SL(2,\zz)$
transformation of the charges and $\Omega$
\be
e^{-i \p\, \bigl(\begin{smallmatrix} \scriptscriptstyle{P}\\ \scriptscriptstyle{Q}\end{smallmatrix} \bigr)^{\!\dag}\! \O
\bigl(\begin{smallmatrix}  \scriptscriptstyle{P}\\ \scriptscriptstyle{Q}\end{smallmatrix} \bigr)}=
e^{-i \p\, \bigl(\begin{smallmatrix} \scriptscriptstyle{P_\g}\\ \scriptscriptstyle{Q_\g}\end{smallmatrix} \bigr)^{\!\dag}\! \O_\g
\bigl(\begin{smallmatrix}  \scriptscriptstyle{P_\g}\\ \scriptscriptstyle{Q_\g}\end{smallmatrix} \bigr)}\;,
\ee
while we also see from the \(Sp(2,\zz)\) transformation property (\ref
{sp2z}) that the modular form is invariant under the transformation of its argument: $\O\to\O_\g$.
Furthermore, using  the fact \(P^2,Q^2\!=\! 0\text{ mod } 2\) and
\(ad\!+\! bc\!=\! 1 \text{ mod } 2\) one  shows that
\be
(-1)^{(P\cdot Q)}\!=\! (-1)^{(P_\gamma\cdot Q_\gamma)}.
\ee
Therefore the integrand of the degeneracy formula (\ref{DVV1}) is invariant under \(S\)-duality.
This fact is not yet sufficient, however,  to prove the invariance of the degeneracies. Namely, due to the presence of the poles,
the expression for $D(P,Q)$ fails to be \(S\)-duality invariant, unless the contour ${\mathcal C}$ is also transformed to a
new contour ${\mathcal C}_\g$. Namely, the equality
\be
\label{Sinvariance}
\oint_{\mathcal{C}} d\O \,
\frac{e^{-i \p\, \bigl(\begin{smallmatrix} \scriptscriptstyle{P}\\ \scriptscriptstyle{Q}\end{smallmatrix} \bigr)^{\!\dag}\! \O
\bigl(\begin{smallmatrix}  \scriptscriptstyle{P}\\ \scriptscriptstyle{Q}\end{smallmatrix} \bigr)}}{\F(\O) } \,
(-1)^{(P\cdot Q)}\
=\
\oint_{\mathcal{C_\g}} d\O_\gamma \, \frac{e^{-i \p\, \bigl(\begin{smallmatrix} \scriptscriptstyle{P_\g}\\
\scriptscriptstyle{Q_\g}\end{smallmatrix} \bigr)^{\!\dag}
\O_\g \bigl(\begin{smallmatrix}  \scriptscriptstyle{P_\g}\\ \scriptscriptstyle{Q_\g}\end{smallmatrix} \bigr)}}{\F(\O_\g)}
\; (-1)^{(P_\g\cdot Q_\g)}
\ee
only holds when upon inserting (\ref{transform_O}) on the r.h.s., the
new contour ${\mathcal C}_\g$ in the $\Omega_\g$-plane is the same as ${\mathcal C}$ in the $\Omega$-plane.

The contours ${\mathcal C}$ and ${\mathcal C}_\g$ are really different, and in general cannot be deformed
into one another without picking up any residue. Therefore, one concludes that under $S$-duality the choice of contour has to change.
A natural way to achieve this is to let the contour depend on the charges, and possibly also the moduli fields, since these quantities do
transform under \(S\)-duality. Indeed, there is an important reason to suspect that the dyon counting formula is moduli-dependent, since it is known that certain  multi-centered BPS solutions only exist in some range of background moduli and decay when a wall of marginal stability
is crossed. The aim of this paper is to determine the charge and moduli dependence of the contour, so that it is consistent with \(S\)-duality and also takes into account the decay of dyonic bound states.

\bigskip

\bigskip

\section{Contour Dependence and Pole Contributions}
\setcounter{equation}{0}

In this section we will examine the contour dependence of the dyon counting formula in details.
For this discussion it will be convenient to parametrize the elements of $\Omega$ explicitly as
\be
\O = \left(\begin{array}{cc} \r & \n \\ \n & \s \end{array}\right)\;.
\ee
In terms of these quantities, the formula for the microscropic dyon degeneracies $D(P,Q)$ takes the form\\[.5mm]
\be
\label{DVV2}
- \oint_{{\mathcal{C}}} d\r d\s d\n\,\frac{e^{-i\pi ({P^2 \rho+Q^2 \s+ (P\cdot Q) (2\n+1))}}}{\Phi(\r,\s,\n)}\;,
\ee

\noindent
where $\mathcal C$ is a contour in the Siegel domain defined by
$$
\im\r>0,\qquad \im\s > 0,  \qquad  \im\r\,\im\s> (\im\n)^2.
$$
In the original proposal of \cite{Dijkgraaf:1996it}, the degeneracies were expressed in terms of the
expansion coefficients of $1/\Phi$ in powers of $e^{2\pi i\r}$, $e^{2\pi i \s}$, and  $\ e^{2\pi i \n}$.
As explained above, this prescription is somewhat ambiguous, since the expansion will
depend on the location of the contour with respect to the poles.

\begin{figure}
\centering
$\begin{array}{ll}
\includegraphics[width=6.2cm]{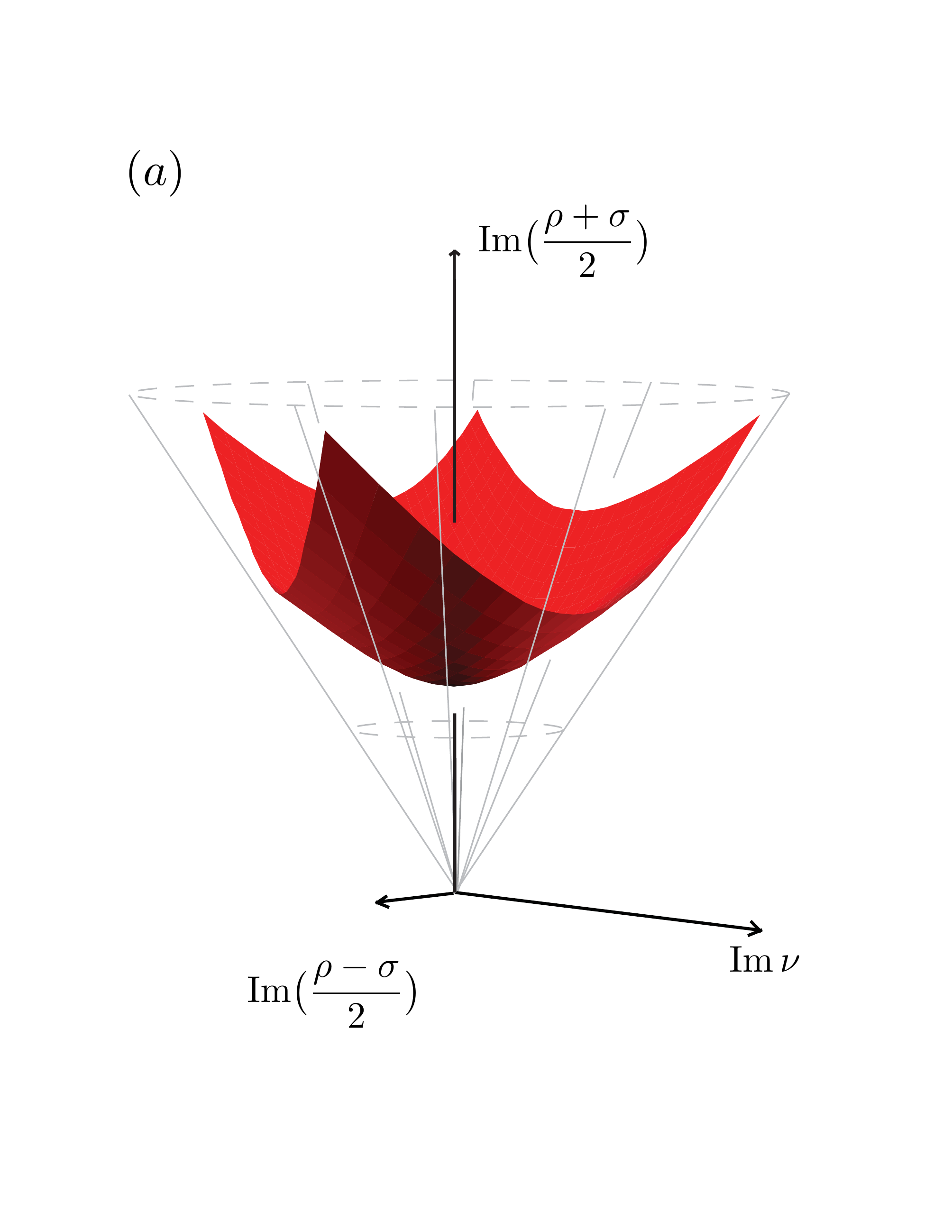} &
\includegraphics[width=6.2cm]{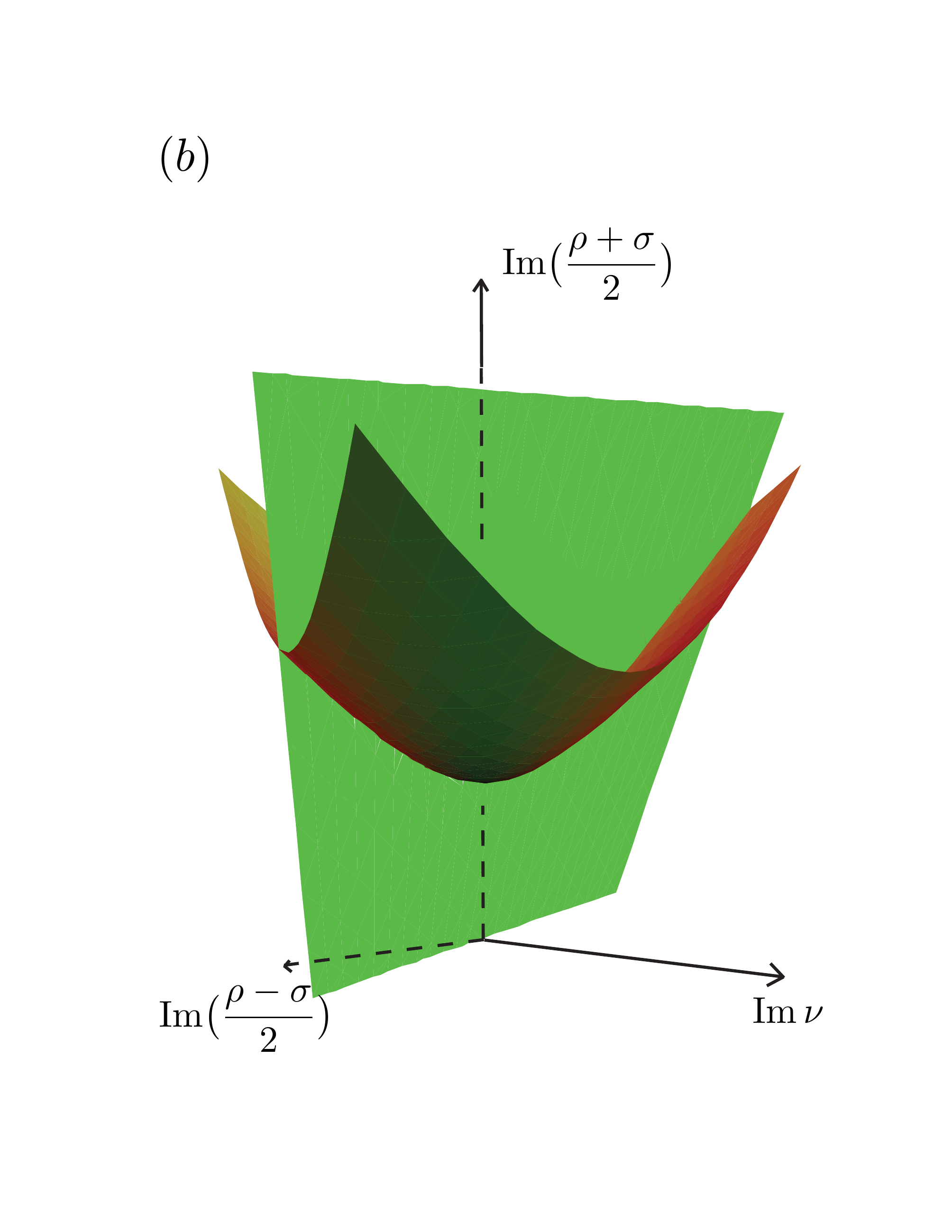}
\end{array}$
\setlength{\abovecaptionskip}{5pt}
\caption{\footnotesize{\label{lightcone} {\bf{(a)}} The Siegel upper-half plane for the modular form \(\F\) is the future light-cone in the
Minkowski space \(\rr^{1,2}\), and we consider the space of all contours to be a sheet of hyperboloid inside this light-cone, with all the points
on the hyperboloid having a large distance from the origin.
{\bf{(b)}} A pole corresponding to an element \(\g\in SL(2,\zz)\) is a plane in \(\rr^{1,2}\) which always intersects the hyperboloid along a
hyperbola. }}
\setlength{\belowcaptionskip}{5pt}
\end{figure}

So let us have a closer look at the possible choice of the contour ${\mathcal{C}}$ in (\ref{DVV2}).
Due to the fact that we are dealing with a modular form, the contour will have to be inside a fundamental domain of the
$Sp(2,\zz)$ modular group. A natural choice of contour is to perform the integral over the real parts of $\r$ $\s$ and $\n$,
while keeping the imaginary parts fixed.
Specifically, the integration range of the real variables is
\be
\label{realdomain}
0\leq \re\r,\re\s,\re\n < 1\;.
\ee
The integration contour is thus a three-torus.
The location of the contour is determined by a choice of the imaginary parts.
To make sure that $1/\Phi$ has a well-defined expansion up to high order, we will choose these imaginary parts so that $\Omega$ lies well inside the Siegel
upper half plane, that is
\be\label{contour_space}
\text{det} (\im \Omega) = \im \r\, \im\s-(\im \n)^2 \ \gg 1\,.
\ee
To visualize the location of the poles relative to the contours, it is convenient to regard the vector
$$
(\im\r,\im\s,\im \n) \in \rr^{1,2}
$$
as a vector in a three dimensional Minkowski space, with $\det(\im\Omega)$ playing the role of the
$SO(1,2)$ invariant inner product. The Siegel domain corresponds to the space inside the future light-cone, while the
space of contours for a given large value of $\det(\im\Omega)$ is identified with a
sheet of a hyperboloid far out inside the future light-cone. This is shown in Figure [\ref{lightcone}].
As mentioned before, all the double poles of the generating function \(1/{\F}\) are located at divisors
given by the \(Sp(2,\zz)\)  modular images of the divisor \(\n=0\)  in the
(\(\r,\s,\n\)) space. The location of this $\nu=0$ divisor can be viewed as a hyperplane which transversely
intersects the space of contours. The location of the other poles can be obtained by acting with the
\(Sp(2,\zz)\) group. Note that the $Sp(2,\zz)$ group action can be identified with that of the conformal group of \(\rr^{1,2}\). This fact can be used to show that these general divisors take the form
\bea \label{poles1} \nonumber
k\r+\ell \s + m \n + r \;(\r\s-\n^2) + s &=& 0 \;\;\;\text{       with} \\
k,\ell,m,r,s \in \zz \;\;\;,\;\;\;m^2 - 4k\ell + 4 rs &=& 1  \;.
\eea
 The poles at divisors with \(r=1\) have exponentially dominant contribution to the degeneracy formula
(\ref{DVV2}) compared to the rest in the case of large charges, as explained in the appendix of
\cite{Dijkgraaf:1996it}.
In \cite{Sen:2007vb} it was observed
that the contour space (\ref{contour_space}) does not intersect any of the
poles having \(|r| \geq1 \). Indeed, a look at the real part of the above equation reveals that, since
all the entries of \(\re\O\) run between 0 and 1, there is nothing to compensate the large contribution from
$\det(\im\Omega)\gg 1$ contained in the real part of $\r\s\!-\!\n^2$.
Hence, these poles will always contribute to the degeneracy formula no matter which contour we
choose, since they lie  lower in the light-cone. Therefore, we
never run into the danger of having a contour which crosses one of these poles.
For our purpose of studying the contour dependence of the integral, it is therefore sufficient to
concentrate on the poles with \(r=0\).

Since we are only interested in the poles inside the real domain of integration (\ref{realdomain}), we can restrict our attention to
the poles with $r\!=\!s\!=\!0$. It is easily seen that these are  the images of the
pole \(\n=0\) under  the \(SL(2,\zz)\) subgroup of \(Sp(2,\zz)\), and hence can be labelled by
the group elements $\gamma$ of \(SL(2,\zz)\)\footnote{One can
show this by, for example, classifying both sets of numbers (\(a,b,c,d\)) and (\(k,\ell,m\)) by their prime
factorizations.}.
Specifically, in terms of the integral matrix elements $a$, $b$, $c$, and $d$ of $\gamma$ we have
$$
k=-bd,\qquad \ell=-ac \qquad m=ad+bc\;,
$$
where the length condition
\be
\label{length}
m^2 - 4k\ell =(ad+bc)^2 - 4abcd =  1
\ee
follows directly from $ad-bc=1$.
The plane inside \(\rr^{1,2}\) defined by the imaginary part of the equation (\ref{poles1}) for this case can thus be written as
\be
\im(-bd\,\r-ac\, \s+(ad+bc)\n) = 0\;.
\ee
The length condition (\ref{length}) implies that the normal vector to the plane is spacelike, and hence
these planes always intersect the contour space hyperboloid (\ref{contour_space}) along a hyperbola. Therefore, each plane
divides the contours into two sub-classes
in a way analogous to the pole \(\n\!=\!0\) (see Figure [\ref{lightcone}]).
Whether the corresponding poles  contribute to the degeneracy formula for a given charge configuration will therefore
depend on the contour we choose.

\begin{figure}
\centering
$\begin{array}{cc}
\includegraphics[width=5.2cm]{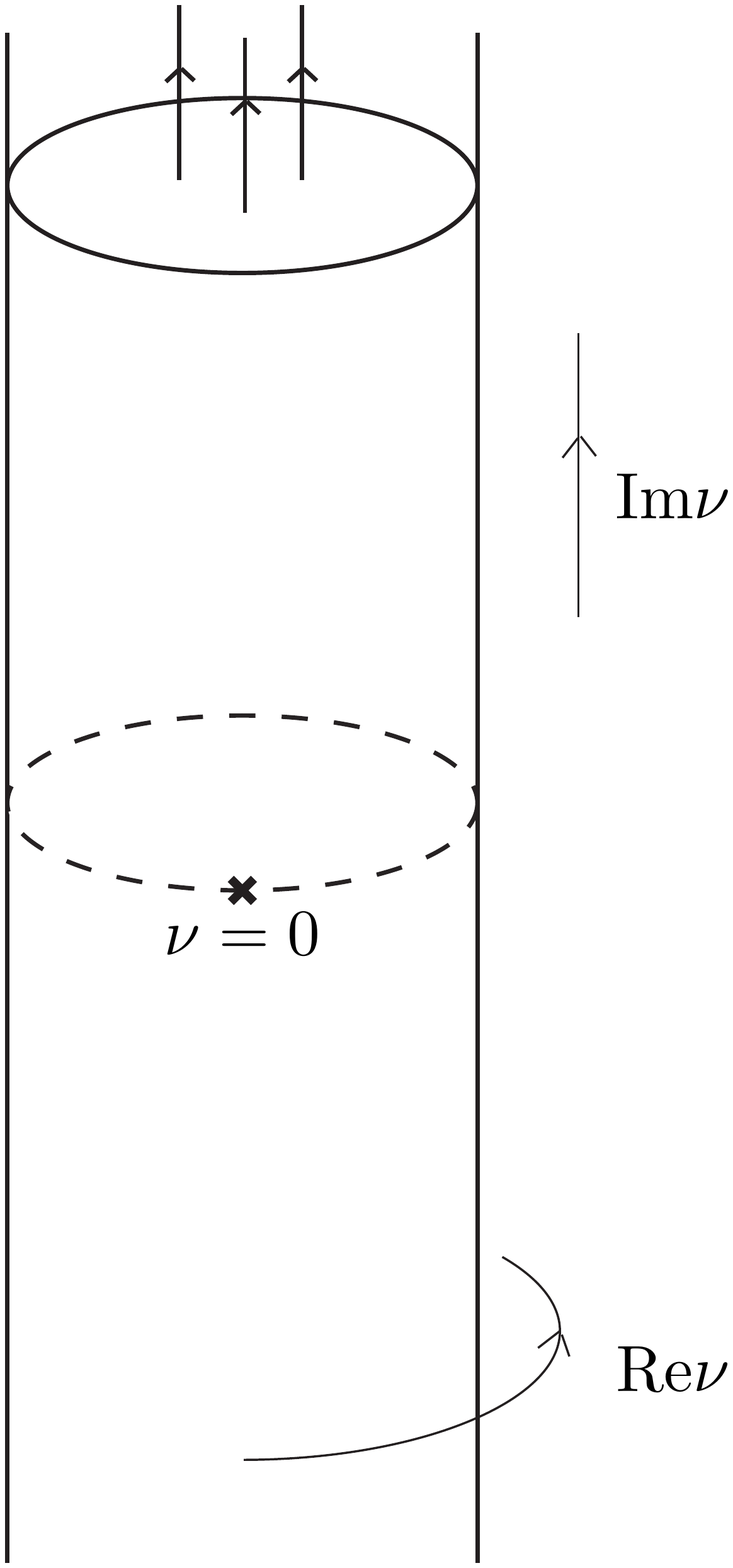} &
\includegraphics[width=5.2cm]{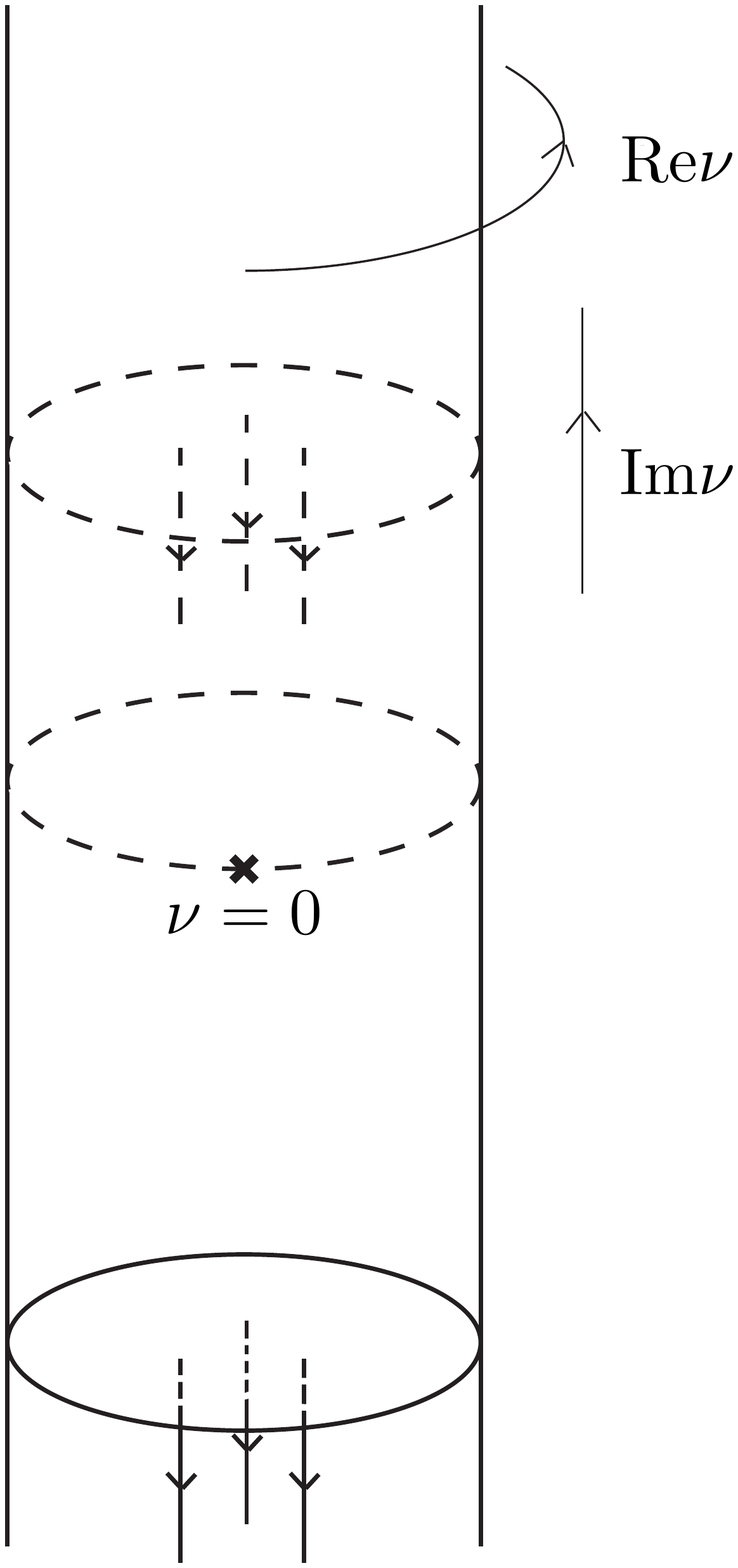}
\end{array}$
\setlength{\abovecaptionskip}{5pt}
\caption{\footnotesize{\label{contour_graph}In this figure we show how the pole located at \(\n=0\)
contributes to the degeneracy formula for contours with \(\im\nu>0\). {\bf{(a)}} For charges with \(P\cdot Q
< 0\), one can deform the contour to the upper infinity of the cylinder where the integrand goes to zero
without hitting the pole. {\bf{(b)}} For charges with \(P\cdot Q > 0\), one can deform the contour to the
lower infinity of the cylinder, and by doing so pick up the residue of the pole. }}
\setlength{\belowcaptionskip}{5pt}
\end{figure}

Let us now determine the condition under which these poles contribute, and, if they do,
calculate their contribution. We first concentrate on  the double pole at $\nu=0$.
Near the $\nu=0$ divisor the generating function has the limit
\be \label{pole1}
\frac{1}{\F(\r,\s,\n)} = -\frac{1}{4 \p^2} \frac{1}{\n^2} \frac{1}{\eta^{24}(\r)} \frac{1}{\eta^{24}(\s)} \;(1+{\cal{O}}
(\n^2))\;.
\ee
Notice that the last two factors in the limiting
expression (\ref{pole1}) are exactly the generating function for the \(1\!/2\)-BPS degeneracies (\ref{half_BPS}).
By plugging the expression (\ref{pole1}) into the degeneracy formula (\ref{DVV2}) and after performing the integration over the real part of $\r$ and $\s$,
one gets
\begin{displaymath}
\frac{(-1)^{P \cdot Q}}{4\p^2} d(P) d(Q)  \oint_{{\mathcal{C}}_\n}  d\n\,\frac{e^{-2 \p i (P\cdot Q) \n}}{\n^2}\;,
\end{displaymath}
where we have made use of (\ref{half_BPS}).
To evaluate the remaining integral over $\nu$, we first consider a contour with \({\rm Im}\n> 0 \).
For this case the contour is shown in the Figure [\ref{contour_graph}].
When the charges under consideration satisfy \(P\cdot Q < 0\), one can deform the contour to the upper infinity
of the cylinder (\(\im\n  \rightarrow \inf \))
where the integrand is zero without crossing any pole. One thus concludes that the integral yields zero.
On the other hand, in the case \(P\cdot Q > 0\), the contour can be moved to the lower infinity (\(\im\n
\rightarrow -\inf \)) where the integrand is again zero, but now by doing so we pick up the contribution of
the pole
\be
-2\p i \pa_{\n} (e^{-2 \p i (P\cdot Q) \n})| _{\n=0} = -4\p^2 \; (P\cdot Q)\;,
\ee
where the extra minus sign comes from the fact that we are enclosing the pole in a clockwise direction.
For the contours with \(\im\nu < 0 \), a similar argument shows that the pole only contributes when $(P\cdot Q)<0$, but now with the opposite sign as above due to the reverse orientation in which the pole is enclosed.  One therefore concludes that the contribution of
this specific pole to the degeneracy formula (\ref{DVV1}) is
\be\label{degeneracy_PQ1}
(-1)^{(P\cdot Q)+1}\, |P\cdot Q|\,d(P)\,d(Q)\;\qquad \mbox{when}\ \ (P \cdot Q)\,\im\nu>0
\ee
and zero otherwise.
The contributions of the other poles can be determined directly in a similar fashion. However, they are more easily
obtained by making use the fact that they are the $SL(2,\zz)$ images of the $\n=0$ pole.
Together with the fact that the integrand is invariant under \(S\)-duality (\ref{Sinvariance}),
it follows that the double pole of \(1/\F\)
located at
\be\label{pole_gamma}
\nu_\gamma\equiv -bd \,\r -ac\,\s +(ad+bc) \n = 0
\ee
gives the contribution
\be \label{jump_gamma}
(-1)^{P_\g\cdot Q_\g+1} \,|P_\g\cdot Q_\g| \,d(P_\g)\,d(Q_\g)\;\qquad \mbox{when}\ \  (P_\g \cdot Q_\g)\,\im\nu_\g>0
\ee
and zero otherwise.
The equation (\ref{jump_gamma}) summarizes all the contour dependence in the degeneracy formula (\ref{DVV2}).

As we will see, the jumps in the counting formula when a contour crosses one of the poles are related to the decay of marginally
bound \(1\!/2\)-BPS particles.
Specifically, we will argue that (\ref{degeneracy_PQ1}) precisely counts the number of states associated with the bound state of
a purely electric \(1\!/2\)-BPS object and a purely magnetic \(1\!/2\)-BPS object, while (\ref{jump_gamma}) is associated with more
general dyonic bound states that are obtained by electric-magnetic duality. This interpretation will be discussed in more
details in section 5, after we describe the supergravity solution corresponding to
these states.

\bigskip

\bigskip

\section{Dying Dyons and Walls of Marginal Stability}
\setcounter{equation}{0}

The central charge in the \({\mathcal{N}}=4\) supersymmetry algebra can be written as
\be \label{central_charge}
\hat{Z}=\frac{1}{\sqrt{\tau_2}}(P_L - \tau Q_L)^m\Gamma_m\mbox{           ;            } m=1,..,6\;,
\ee
where $\tau=\tau_1+i\tau_2$ is the usual complexified axion-dilaton field
and the left-moving charges
are given by a six-dimensional projection of the 28 dimensional charge vectors,
\be \label{projection}
P_L =  \m^m_A\, P^A ,\qquad Q_L = \m^m_A\, Q^A\mbox{             ;           }A=1,2,...., 28\;.
\ee
Here \(\m^m_A\) is a \(6\times28\) matrix comprised of the \(6\times 22\)
moduli fields parametrizing the coset space \( \frac{O(6,22)}{O(6)\times O(22)} \).
Here and from now on all the moduli fields are evaluated at spatial infinity.

The square of the BPS mass is equal to the largest eigenvalue of $\hat{Z}^\dagger \hat{Z}$. One can
choose the basis such that all the gamma matrices are hermitian, and one then finds
\be
\hat{Z}^\dagger \hat{Z}=\frac{1}{\tau_2}| P_L- \tau Q_L|^2\,{\mathds {1}} -2i  P_L^m Q_L^n \Gamma_
{mn}\;.
 \ee
 From the fact that the operator \( i P_L^m Q_L^n \Gamma_{mn}\) satisfies
 \be
 (i P_L^m Q_L^n \Gamma_{mn})^{2} =   |P_L\wedge Q_L|^2 \equiv Q_L^2 \,P_L^2 - (Q_L\cdot P_L)^2\;,
 \ee
one concludes that $\hat{Z}^\dagger \hat{Z}$ has the following two eigenvalues
\bea
 |Z_{P,Q}|^2&=& {1\over \tau_2}| P_L-\tau Q_L|^2+2|P_L\wedge Q_L|\\  \text{and            }\;\;\;\;\;\;\;
|Z'_{P,Q}|^2& =&{1\over \tau_2}| P_L-\tau Q_L|^2-2|P_L\wedge Q_L|\;.
\eea
The complex number $Z_{P,Q}$ with the largest norm plays in theories with ${\cal N}=4$ supersymmetry the same
role as the single central charge in ${\cal N}=2$ theories. In particular, it determines the
BPS mass
$$
M_{P,Q}=|Z_{P,Q}|
$$
and will therefore simply be referred to as the central charge.
By choosing a specific spinor basis one can also fix the phase of \(Z_{P,Q}\). We denote it by
\(\a_{P,Q}\), {\it{i.e.}}
\be
Z_{P,Q} = e^{i\a_{P,Q}} \, |Z_{P,Q}|\;.
\ee
Note, however, that this phase is not unambiguously defined, since it depends on the choice of the spinor basis.
As we will explain later, the criterion that determines whether two \(1\!/2\)-BPS
objects form a bound state can be formulated as a condition on the relative phase between the
central charges of different objects. This relative phase is
independent of the choice of spinor basis, even though the
overall phase of the central charges is not.

A related comment is the following. Consider the \(SL(2,\zz)\) duality
transformation of the charges  (\ref{sl2z_charge}) and the axion-dilaton moduli (\ref{sl2z_tau}),
this transformation has the
effect of shifting the phase of the central charges by
\be
\label{central_charge_transform}
Z_{P,Q} \to e^{-i\alpha_\g} Z_{P,Q},\qquad \alpha_\g= \arg(c\t+d)\;.
\ee
Again, the phase shift is independent of the charges, therefore all the relative phases will indeed be duality invariant.

We are now interested in knowing when a dyonic bound state might decay. First we concentrate on the
specific decay channel of a dyonic, \(1\!/4\)-BPS state with charges \((P,Q)\) splitting into two \(1\!/2\)-BPS
particles with charges \((P,0)\) and \((0,Q)\).
For this case, the condition for a wall of  marginal stability is
\be \label{mass}
M_{P,Q}=M_{P,0}+M_{0,Q}\;,
\ee
which can be rewritten as
\be
|Z_{P,Q}| = |Z_{P,0}| + |Z_{0,Q}| \;.
\ee
Using the fact that the total central charge obeys
\be
Z_{P,Q}=Z_{P,0}+Z_{0,Q}\;,
\ee
one finds that the condition of marginal stability can only be satisfied when
the phases of the central charges are aligned. An explicit expression for this
condition can be obtained either by determining these phases, or directly from (\ref{mass}) by using the explicit formula for the
BPS mass. Both approaches require a little bit of manipulation, and lead to the condition
\be \label{line1}
{\tau_1\over \tau_2}+{{P_L\cdot Q_L}\over |P_L\wedge Q_L|} =0\;.
\ee
The next step will be to consider the other ways in which a dyon can split into two \(1\!/2\)-BPS particles, and determine the
corresponding walls of marginal stability.
By  definition a \(1\!/2\)-BPS state must have degenerate eigenvalues of the operator \(\hat{Z}^\dag\hat{Z}\) and
thus have parallel electric and magnetic charges
\be
|Z_{P,Q}|^2 = |Z'_{P,Q}|^2 \Leftrightarrow P \parallel Q\;.
\ee
As discussed in \cite{Sen:2007vb}, these \(1\!/2\)-BPS decay channels can be labelled by \(SL(2,\zz)\)
elements as\footnote{Strictly speaking, the bound states are labeled by elements of \(PSL(2,\zz)\),
since states related by exchanging the two particle are physically equivalent.}
\be\label{decay1}
 \Bigl( \begin{array}{c}P\\ Q \end{array} \Bigr)
=  \g^{-1}  \Bigl(\begin{array}{c}P_\g\\ 0 \end{array}\Bigr)  + \g^{-1} \Bigl(\begin{array}{c} 0 \\ Q_\g
\end{array}\Bigr)
\equiv  \Bigl(\begin{array}{c}  P_1 \\ Q_1 \end{array}\Bigr) + \Bigl(\begin{array}{c}  P_2 \\ Q_2 \end{array}\Bigr) \;,
 \ee
where the two terms to the right of the equivalence sign are defined by the corresponding terms left of this sign.
This equation shows that these bound states are basically the $SL(2,\zz)$ transforms of the bound state of a purely electric particle with charge
$P_\g$ and a magnetic particle with charge $Q_\g$. Using the fact that the central charge $Z_{P,Q}$ is $SL(2,\zz)$ invariant up to a phase,
one finds  the condition of marginal stability to be
\be
|Z_{P,Q}|_\tau =|Z_{P_\g,0}+Z_{0,Q_\g}|_{\tau_\gamma}=|Z_{P_\g,0}|_{\tau_\gamma}+|Z_{0,Q_\g}|_{\tau_\gamma}\;,
\ee
where in the last two expressions the central charge is evaluated with the $SL(2,\zz)$ transformed value of the axion-dilaton fields
\be
\label{sl2z_tau}\t_\g \equiv \frac{a\t+b}{c\t+d}\;.
\ee
After a straightforward calculation, the position of the corresponding wall of marginal stability turns out to be precisely given by the $SL(2,\zz)$ image of the one for the bound state of the
purely electric and purely magnetic \(1\!/2\)-BPS states. Namely, the walls of marginal stability for all two-centered \(1\!/2\)-BPS splits are
\be
\label{wallgamma}
\frac{\t_{\g,1}}{\t_{\g,2}}+ \frac{(P_{L} \cdot Q_{L})_\g}{|P_{L} \wedge Q_{L} |_\g}
=0\;.
\ee
As mentioned in \cite{Sen:2007vb}, the projection of the above wall of
stability from the full 134 dimensional moduli space to the upper \(\t_\g\)-plane is a straight line. But when regarded
in the original \(\t\)-plane, it is a circle for generic group
elements \(\g\).

The meaning of the presence of a wall of marginal stability is that a BPS bound state of two particles
exists on one side of the wall and disappears when crossing into the other side. After deriving the location
of the walls for these bound states, we would like to know on which side these states are stable
and on which side unstable. For this purpose we need more information about the corresponding supergravity solutions.

\bigskip

\begin{flushleft}
{\it{Stability Conditions from the Supergravity Solutions}}
\end{flushleft}

\noindent
Let us now consider the four-dimensional \({\cal{N}}= 4\) supergravity theory describing the low energy
limit of the heterotic string compactified on a six-torus.
The metric part of a stationary solution reads
\bea
ds^2&=& -e^{-2U} (dt + \vec{\o}\cdot d\vec{x})^2 + e^{2U}d\vec{x}^2\\
e^{2U}&=&{ |\cal{P}\wedge \cal{Q}|} \,{\equiv}
\sqrt{  {\mathcal{P}}^{2} {\mathcal{Q}}^2 - ( {\mathcal{P}} \cdot {\mathcal{Q}})^2 } \\
\label{angular_momentum1}
\vec{\nabla} \times \vec{\o} &=& \cal{P}\cdot \vec{\nabla} \cal{Q}- \cal{Q}\cdot \vec{\nabla} \cal{P} \;,
\eea
where the indices are contracted using the standard \(SO(6,22)\)-invariant \(28\times 28\) matrix \(\eta_{AB}\), for
example \({\cal{P}}^2 \equiv {\cal{P}}^A {\cal{P}}^B \eta_{AB}\).

The 56 harmonic functions appearing in the above solution are
\bea\nonumber
{\cal{P}}^A(\vec{x}) = C^A + \sum_i \frac{P^A_i}{|\vec{x}-\vec{x}_i|}&\\
{\cal{Q}}_A(\vec{x}) = D_A + \sum_i \frac{Q_{A,i}}{|\vec{x}-\vec{x}_i|}&\;,
\eea
with the 56 constants given by the asymptotic value of 23 complex scalar fields (the axion-dilaton moduli
\(\t\) and the 22 complex moduli projected from the aforementioned \(6\times 22\) moduli) as\footnote
{By evaluating the \({\mathcal{N}}=4\) central charge operator \(\hat{Z}\) (\ref
{central_charge}) in the eigen basis of  \(\hat{Z}^\dagger\hat{Z}\), one
can write the BPS equations in a way analogous to the \({\mathcal{N}}=2\) case \cite{Denef:2000nb}.
Only 22 complex moduli made out of the \(6\times 22\) real moduli fields play a role in the solution.
It is indeed known that the \({\mathcal{N}}=4\) moduli space locally decomposes as a product of 22 vector-, 44 hyper-,
and 1 tensor-multiplet scalars in the \({\mathcal{N}}=2\)  language (see, for example, \cite{Moore:1998pn}).}
\bea\nonumber
C^A =-   \im\left(e^{-i\a_{P,Q}} \frac{\pa Z_{P,Q}}{\pa Q_A} \right) &\\
D_A =   \im\left(e^{-i\a_{P,Q}} \frac{\pa Z_{P,Q}}{\pa P^A} \right) &\;,
\eea
where the \(P^A\)'s and the \(Q_A\)'s denote the total charges coming from all the centers.
From this expression one immediately sees that these coefficients satisfy \( Q_A C^A = P^A D_A \),
since the central charge is linear in all charges.

For the specific two-center bound state considered earlier, the corresponding supergravity solution
has harmonic functions given by
\bea\nonumber
{\cal{P}}^A = C^A +  \frac{P^A}{|\vec{x}-\vec{x}_P|}&\\
{\cal{Q}}_A = D_A +  \frac{Q_{A}}{|\vec{x}-\vec{x}_Q|}&\;.
\eea
In this case the coordinate distance between the two centers \(|\vec{x}_P-\vec{x}_Q|\) is fixed by the
integrability condition \cite{Denef:2000nb}, obtained by taking the divergence of the both sides of (\ref{angular_momentum1}), and reads
\be
\frac{P\cdot Q }{|\vec{x}_P-\vec{x}_Q| } = - C^A Q_A\;.
\ee

After some algebra this becomes
\be
\frac{{P\cdot Q}}{|\vec{x}_P-\vec{x}_Q|} = - \frac{{|P_L\wedge Q_L|}}{M_{P,Q}} \,\left(  \frac{\t_1}{\t_2}+\frac{P_L \cdot
Q_L}{|P_L \wedge Q_L |}
 \right)\  \;.
\ee
Since the distance between the two centers is always a positive number, one finds that, in order for the bound state to exist,
the expression on the r.h.s. must have the same sign as $P\cdot Q$.
Therefore the bound state only exists when
\be\label{stability_1}
-(P\cdot Q)\, \left( \frac{\t_1}{\t_2}+\frac{P_L \cdot Q_L}{|P_L \wedge Q_L |}
 \right)\ > 0\;,
\ee
and decays when one tunes the background moduli to hit the wall where the above expression vanishes.
More precisely,  one finds that the distance between the two centers goes to infinity,
and the bound state no longer exists as a localizable state.

For the other bound states of \(1\!/2\)-BPS particles obtained by
acting with an element $\gamma$ of the electric-magnetic duality group there exist similar solutions. But in this case the harmonic functions will have
a seemingly more complicated form than the \(P|Q\) split studied above. More explicitly, now the
harmonic functions and the corresponding integrability condition takes the form
\bea\nonumber
{\cal{P}}^A = C^A +  \frac{P_1^A}{|\vec{x}-\vec{x}_{P_\g}|}+\frac{P_2^{A}}{|\vec{x}-\vec{x}_{Q_\g}|}&\\
{\cal{Q}}_A = D_A +  \frac{Q_{A,1}}{|\vec{x}-\vec{x}_{P_\g}|}+ \frac{Q_{A,2}}{|\vec{x}-\vec{x}_{Q_\g}|}&\;,
\eea
and
\be
\left(\frac{Q_2}{|\vec{x}_{P_\g}-\vec{x}_{Q_\g}|} + D\right)\cdot P_1 -
\left(\frac{P_2}{|\vec{x}_{P_\g}-\vec{x}_{Q_\g}|} + C\right)\cdot Q_1 = 0\;,
\ee
where \(P_{1,2}\), \(Q_{1,2}\) are given in terms of the original charges and the group element \(\g\) as
(\ref{decay1}). Plugging in the charges, and after some manipulations using (\ref{central_charge_transform}),
the above integrability can be written, as expected, in a similar form as above:
\be
\frac{(P_\g \cdot Q_\g)}{|\vec{x}_{P_\g}-\vec{x}_{Q_\g}|} =- \frac{{|P_L\wedge Q_L|}}{M_{P,Q}}
\left( \frac{\t_{\g,1}}{\t_{\g,2}}+\frac{(P_{L} \cdot Q_{L})_\g}{|P_{L} \wedge Q_{L} |_\g}
 \right)\;.
\ee
Thus, following the same reasoning,  one finds exactly the same stability condition
\be
\label{stability_2}
-(P_\g\cdot Q_\g)\, \left( \frac{\t_{\g,1}}{\t_{\g,2}}+\frac{(P_{L} \cdot Q_{L})_\g}{|P_{L} \wedge Q_{L} |_\g}
 \right)\  > 0\;,
\ee
but now with both the charges and the axion-dilaton transformed with \(\g\in SL(2,\zz)\).

\bigskip

\bigskip

\section{The Contour Prescriptions and Their Interpretation}
\setcounter{equation}{0}

Let us now return to the problem of identifying the contour that should be used in the counting formula, so that it counts the right number of states for a given
value of the moduli. The key observation which will allow us to find the right prescription is that the contour dependence due to the crossing of
 the pole labelled by $\gamma$ should exactly match the physical decay process of the corresponding dyonic bound state. For example, at the wall
of marginal stability of the bound state of an electric \(1\!/2\)-BPS particle with charge $P$ and a magnetic
\(1\!/2\)-BPS particle with charge $Q$, one expects the degeneracy $D(P,Q)$ to be adjusted by a certain amount
corresponding to the degeneracy of this \((P,0)\), \((0,Q)\) bound state. This degeneracy can be found in the following way
\cite{Sen:2007vb,Dabholkar:2007vk,Denef:2007vg}.
Firstly, each of the two centers has its respective degeneracy \(d(P)\), \(d(Q)\), which is given by the \(1\!/2\)-BPS partition function of the theory as (\ref{half_BPS}).
Secondly, there is an extra interaction factor due to the fact that the spacetime is no longer static.
The conserved angular  momentum, after carefully quantizing the system \cite{Denef:2002ru},  turns out to be
\be\label{angular_momentum}
2J+1= |P\cdot Q|\;.
\ee
One therefore concludes that the jump in the counting formula when one crosses the wall of marginal stability from the stable to the unstable side
is given by
\be
D(P,Q)\to D(P,Q) + (-1)^{(P\cdot Q)}\, |P\cdot Q|\,d(P)\,d(Q)\;.
\ee
This jump in the degeneracy is precisely the contribution (\ref{degeneracy_PQ1}) that we found from the pole at $\nu=0$!
Similar jumps occur when one crosses the walls of marginal stability for the other dyonic states labelled by
$SL(2,\zz)$ elements $\gamma$.
These jumps are again precisely given by the contributions (\ref{jump_gamma}) of the poles at $\nu_\gamma=0$.
In terms of the contour space parametrized by $\im\r$, $\im\s$ and $\im \nu$,
we have shown that whether this pole contributes or not depends on the sign of $(P_\g\cdot Q_\g)\,\im\nu_\gamma$,
while from the supergravity solution we have learned that whether this bound state
exists or not depends on the sign of the l.h.s. of (\ref{stability_2}).
It is therefore natural to make the identification
\be
\label{nugamma}
\im\nu_\gamma = -\Lambda \left( \frac{\t_{\g,1}}{\t_{\g,2}}
+\frac{(P_{L} \cdot Q_{L})_\g}{|P_{L} \wedge Q_{L} |_\g}\right)\;,
\ee
where $\Lambda$ is a yet undetermined positive parameter.
This equality actually constitutes an infinite number of equations, namely one for each element $\g\in SL(2,\zz)$.
It is not immediately clear that all these equations can be imposed without running into contradictions.

One way to show their mutual consistency is to work out both sides of the equation, and observe that
the left as well as the right can be written as a sum of products of the integers
$a$, $b$, $c$ and $d$ as in (\ref{pole_gamma}).
By identifying the various terms one arrives at a prescription that is independent of these integers.
In other words, in this way one finds that the infinite set of equations (\ref{nugamma}) are equivalent to the
following three conditions
\be
\label{imnu}
\im\nu= -\Lambda \left( \frac{\t_1}{\t_2}+\frac{P_L \cdot Q_L}{|P_L \wedge Q_L |}
 \right)\;,
\ee
\be
\im\r = \Lambda \left( \frac{1}{\t_{2}}
+\frac{Q_L\cdot Q_L}{|P_{L} \wedge Q_{L} |}\right)\;,
\ee
and
\be
\im\s = \Lambda \left( \frac{|\t|^2}{\t_{2}}
 +\frac{P_L\cdot P_L}{|P_{L} \wedge Q_{L} |} \right)\;.
\ee
These equations determine the location of the contour $\mathcal C$ in terms of the charges and moduli.
To see that these equations are consistent with \(S\)-duality invariance, we better use a
 more clever way to write them. It will turn out to be convenient to introduce
 the \(2\times 2\) matrices
\be
 \label{def_Mtau}
{\cal M}_{\tau}=
\frac{1}{\t_2} \left(\begin{array}{cc} 1 & -\t_1 \\ -\t_1 &|\t|^2 \end{array}\right)
\ee
and
\be \label{def_MPL}
{\cal M}_{P_L,Q_L} \equiv \frac{1}{|P_L\wedge Q_L|} \left(\begin{array}{cc} Q_L\cdot Q_L &  - P_L\cdot Q_L \\ -P_L\cdot
Q_L &P_L\cdot P_L \end{array}\right)\;.
\ee
Notice that the first matrix is given in terms of the asymptotic value of the
axion-dilaton moduli, while the second depends on the charges and
contains the asymptotic Narain moduli.
These matrices transform in an identical fashion under the electric-magnetic \(S\)-duality group, namely
\be
{\cal M}_{\t}\to \left(\gamma^T\right)^{-1}\! {\cal M}_{\t} \gamma^{-1} \qquad\qquad {\cal M}_{P_L,Q_L}\to
\left(\gamma^T\right)^{-1}\!{\cal M}_{P_L,Q_L}\gamma^{-1} \;.
\ee
It is important to note that these transformation rules are the same as those of $\Omega$.
We can now summarize the results of the previous section in terms of these matrices as follows.
The location of the wall of marginal stability (\ref{wallgamma}) labelled by the $SL(2,\zz)$ element $\gamma$
is given by the condition
\be
\label{SLwalls}
\Bigl(\left(\gamma^T\right)^{-1}\! \left( {\cal M}_\tau+{\cal M}_{P_L,Q_L}\right)\gamma^{-1}\Bigr)^\sharp= 0\;.
\ee
Here, the superscript $\sharp$ again denotes the off-diagonal component
of the $2\times 2$ matrix inside the brackets.
Similarly, the location of the corresponding pole $\nu_\gamma=0$ is given in terms of $\Omega$ by
\be
\label{SLpoles2}
\Bigl( \left(\gamma^T\right)^{-1}\!\Omega \gamma^{-1}\Bigr)^\sharp\ = 0\;.
\ee
In this way we are naturally led to the following moduli-dependent contour prescription.
The contour is determined by specifying the value for the
imaginary part of $\Omega$ in terms of the matrices ${\mathcal M}_\t$ and ${\cal M}_{P_L,Q_L}$  containing the background moduli.
The prescription reads
\be
\label{contour_summary}
\im\Omega=\Lambda\left({\cal M}_\tau+{\cal M}_{P_L,Q_L}\right)
\ee
with \(\Lambda\gg1\). Here $\Lambda$ is taken to be large to ensure that the series expansion of
$1/ \Phi$ converges rapidly. Moreover, as explained earlier, for large $\Lambda$ the contour avoids all other poles
except the ones given by $\nu_\gamma=0$.
Also note that the identification (\ref{contour_summary})
is consistent with the Siegel condition, since
\be
\text{det}(\im\O) = \Lambda^2 \left(\frac{M_{P,Q}^2}{|P_L\wedge Q_L|}\right)\Bigr\rvert_{\inf} >0 \;
\ee
and the trace of $\im\Omega$ is also easily seen to have the required sign.
Using the results (\ref{SLwalls}) and (\ref{SLpoles2}),
one easily verifies that this contour precisely crosses the right poles at the walls of marginal
stability to account for the correct jumps in the dyon degeneracies.
Furthermore, note that the contour prescription leads to a manifestly \(S\)-duality invariant counting formula.
\bigskip

\begin{flushleft}
{\it{The Attractor Contour for Large Charges}}
\end{flushleft}

\noindent
For large charges corresponding to a macroscopic black hole, it is natural to ask what happens to our prescription when one takes the
moduli at infinity to be at the attractor value. Since the attractor values of the moduli are completely determined by the charges, this procedure
leads to a degeneracy formula that is independent of the moduli. At the attractor point in moduli space the following equations hold for the Narain moduli
\be
\label{attractor}
P_R|_{\text{\tiny{attr.}}}=0,\qquad Q_R|_{\text{\tiny{attr.}}}=0\;,
\ee
and the axion and dilaton are given by
\be
\label{attractor_1}
\tau_1|_{\text{\tiny{attr.}}} = { P\cdot Q\over Q^2},\qquad \tau_2|_{\text{\tiny{attr.}}}={|P\wedge Q|\over Q^2}\;.
\ee
From these equations  the attractor values of the matrices ${\cal M}_\tau$ and ${\cal M}_{P_L,Q_L}$ are easily determined.
One finds
\be
{\cal M}_{\tau}|_{\text{\tiny{attr.}}}={\cal M}_{P_L,Q_L}|_{\text{\tiny{attr.}}}={\cal M}_{P,Q}\;,
\ee
where the $2\times 2$ matrix $M_{P,Q}$ is defined by
\be \label{def_MPQ}
{\cal M}_{P,Q} \equiv \frac{1}{|P\wedge Q|} \left(\begin{array}{cc} Q\cdot Q &  - P\cdot Q \\ -P\cdot
Q &P\cdot P \end{array}\right)\Bigr\rvert_{\inf}\;.
\ee
Here the inner products between the charges are again defined using the moduli-independent $SO(6,22)$ invariant metric.
In this way, we find that at the attractor point our moduli-dependent contour reduces to the following moduli-independent expression
\be
\label{proposal_2}
\im\O= 2\Lambda {\cal M}_{P,Q}\;.
\ee
Again the $SL(2,\zz)$ invariance is manifest, since both sides transform in the same way, and hence this prescription
also leads to a \(S\)-duality invariant counting formula. But what are the states
that are being counted by this formula? In fact, we will now argue that these are precisely the \(1\!/4\)-BPS states that are not given by the bound states of two
\(1\!/2\)-BPS particles, and therefore cannot decay.
Namely, when one fixes the moduli to be at the attractor values, the stability condition (\ref{stability_1}) reduces to
$$
-2 \frac{(P\cdot Q)_\g^2}{|P\wedge Q|} >0\;,
$$
which can clearly never be satisfied.
In other words, none of the bound states of two \(1\!/2\)-BPS particles can exist at the attractor moduli,
which is a fact consistent with the general phenomenon that  an attractor flow always flows from the stable to the unstable side.
In this sense, our moduli-independent contour prescription leads to a counting formula which counts only the ``immortal" dyonic states that exist everywhere in the
moduli space.
Notice further that this class of contours is not defined for
charges with negative discriminant, since they lie outside of the Siegel domain. Furthermore, they do not have an attractor point,
and there is no single-centered supergravity solution carrying these charges.

Finally we would like to briefly comment on the role of the number \(\L\) in our proposed contours
(\ref{contour_summary}), (\ref{proposal_2}).
It can be seen as playing the role of a regulator for the convergence of the generating function. To see
this, notice that when we take the contour according to our prescription (\ref{proposal_2}), the contribution
\be
\Bigl\lvert \, D(P,Q) \, e^{i \p\, \bigl(\begin{smallmatrix} \scriptscriptstyle{P}\\ \scriptscriptstyle{Q}\end{smallmatrix} \bigr)^{\!\dag}\!
\O\bigl(\begin{smallmatrix}  \scriptscriptstyle{P}\\ \scriptscriptstyle{Q}\end{smallmatrix} \bigr) }  \Bigr\rvert =  |D(P,Q)| e^{-4 \L \p |P\wedge Q|}  \sim e^S e^{-4\Lambda \,S}\;
\ee
of certain large charges to the partition function is highly suppressed when  \(\L\gg1\), and we are therefore left with  a rapidly converging generating function.

\bigskip

\bigskip

\section{Conclusion and Discussion}
\setcounter{equation}{0}

In this paper we establish the precise relation between the contour dependence of the microscopic formula and
the presence/absence of bound states of two \(1\!/2\)-BPS configurations in different parts of moduli space in
the macroscopic supergravity theory. Furthermore we propose a  moduli-dependent prescription for the integration
contour, such that all these two-centered bound states are correctly counted by the counting formula. Therefore we arrive
at the surprising and somewhat unexpected conclusion that the counting formula actually counts the degeneracies in {\it{all }}
 of the moduli space, and that there is a well-defined way to extract these degeneracies from the counting formula by choosing
 the contour appropriately. In particular, the counting formula has a built-in \(S\)-duality invariance when the prescribed contour is used.
 Furthermore, for large black hole charges, we also propose a second, moduli-independent contour by going to the attractor value of the moduli,
 using which only the ``immortal dyons" which exist everywhere in the moduli space are counted.

But there are certainly things we do not yet understand about this counting formula. First of all, what are the meaning of the o
ther poles which seem always to contribute? Poles with \( r>1\) in (\ref{poles1}) give a correction to the index of order
\( e^{S/r}= e^{\p |P\wedge Q|/r}\) for large charges, which suggests that they might account for a split of charges into
\(r\) pieces in some way. It would be nice to understand better the role of other splittings of charges in the BPS spectrum.
A second but not unrelated question is, what are the spacetime interpretation of the modular group \(Sp(2,\zz)\)?
Resorting to the product expression for the generating function \cite{Dijkgraaf:1996it}
\be
\frac{1}{\F(\O)} = \frac{1}{e^{2\p i (\r+\s+\n)}}\,\prod_{(k,\ell,m)>0}\left(\frac{1}{1-e^{2\p i (k\r+\ell\s + m\n)}}\right)^{c(4k\ell-m^2)}
\ee
reveals that all the poles susceptible to contour dependence are caused by the lowest-lying oscillators with multiplicity
\(c(-1) = 2\) (recall that \(c(n)=0\) for \(n<-1\)), and vice versa. This suggests that there might be a way of rewriting the
generating function analogous to the sum over modular images of the polar terms as in \cite{Dijkgraaf:2000fq,Kraus:2006nb,deBoer:2006vg,Denef:2007vg}.
We hope to put these puzzles together and return to these issues in the future.

Another question to be asked is, how would the inclusion of higher-order corrections to the low energy effective action affect our macroscopic analysis?
In which way does, if it does, the counting formula encode the information about these corrections?

Finally, this inverse of the modular form \(\frac{1}{\F(\O)}\) seems to our knowledge to be the first example of a moduli-independent partition
function, in the sense that the index is always summarized by the same generating function, but different expansion points must be used for different
background moduli.  This fact might shed some light on the enigma of the split state counting \cite{Denef:2007vg}, which arises due to the presence of
marginal stability walls in a \({\mathcal{N}}=2\) setup. It would be very interesting to investigate whether some of the similar structure is also present
in (some classes of) \({\mathcal{N}} = 2\), D=4 theories.

As for the CHL models, a dyon counting formula has been proposed  for appropriate \(\zz_N\) orbifolds of the above theory for \(N=2,3,5,7\).
In these theories, the rank of the gauge group is reduced and the \(S\)-duality group is now the following subgroup of \(SL(2,\zz)\):
\be
\G_1(N) = \left\{\bem a& b\\ c&d\eem\in SL(2,\zz)\;;\; c= 0 \text{ mod } N\;, a,d =  1 \text{ mod } N\right\}\;.
\ee
Moreover, the family of the contour-dependent poles of the proposed generating function \(\frac{1}{\til{\F}_k(\til{\O})}\),
which is now a modular form of a subgroup of \(Sp(2,\zz)\), and the ways in which a dyon can split into two \(1\!/2\)-BPS particles,
are both modified compared to the original theory. Nevertheless, we find that they can again both be labelled by the elements of the
\(S\)-duality group \(\G_1(N)\),
and these poles again give the same jump of index as the decaying of these bound states.
In particular, following the same arguments we make exactly the same proposal (\ref{contour_summary})
for the integration contour for the dyon counting formula of this class of models.

\section*{Acknowledgments}

We would like to thank Atish Dabholkar and especially Frederik Denef for useful discussions.
M.C. would like to thank the Galileo Galilei Institute for Theoretical
Physics for the hospitality and the INFN for partial support during the
completion of this work.
This research is supported financially by the Foundation of Fundamental Research on Matter (FOM).

\end{document}